\documentclass[preprint,aps,prd,12point,nofootinbib,showpacs,superscriptaddress]{revtex4-1}

\def\theequation{\arabic{section}.\arabic{equation}}

\usepackage{psfrag} 
\usepackage{subfigure} 
\usepackage{color} 
\usepackage{mathrsfs} 
\usepackage{graphicx} 
\usepackage{amssymb, bm} 
\usepackage{amsmath, amsthm} 
\usepackage{epstopdf} 
\usepackage{hyperref} 
\usepackage{enumerate} 
\usepackage{longtable} 

\usepackage{amsfonts} 

\newcommand\e{\mathrm{e}} 
\newcommand\nn{\nonumber \\} 
\newcommand{\be}{\begin{equation}} 
\newcommand{\ee}{\end{equation}}

\begin{document} \def\theequation{\arabic{section}.\arabic{equation}}

\title{New entropies, black holes, and holographic dark energy}



\author{Shin'ichi~Nojiri}
\email[nojiri@phys.nagoya-u.ac.jp]{}
\affiliation{Department of Physics, Nagoya University, Nagoya 464-8602, 
Japan} 
\affiliation{Kobayashi-Maskawa Institute for the Origin of 
Particles and the Universe, Nagoya University, Nagoya 464-8602, Japan}

\author{Sergei~D.~Odintsov}
\email[odintsov@ice.cat]{}
\affiliation{Instituci\'{o} Catalana de Recerca i Estudis Avan\c{c}ats 
(ICREA), Passeig Llu\'{i}s Companys, 23, 08010 Barcelona, Spain} 
\affiliation{Institute of Space Sciences (ICE-CSIC), C. Can Magrans s/n, 
08093 Barcelona, Spain} 

\author{Valerio Faraoni}
\email[vfaraoni@ubishops.ca]{}
\affiliation{Department of Physics \& Astronomy, Bishop's University, 2600 
College Street, Sherbrooke, Qu\'ebec, Canada J1M~1Z7}

\bigskip \bigskip \begin{abstract}

The Bekenstein-Hawking entropy is a cornerstone of horizon thermodynamics 
but quantum effects correct it, while inequivalent entropies arise also in 
non-extensive thermodynamics. Reviewing our previous work, we advocate for 
a new entropy construct that comprises recent and older proposals and 
satisfies four minimal key properties. The new proposal is then applied to 
black holes and to holographic dark energy and shown to have the potential 
to cause early universe inflation or to alleviate the current Hubble 
tension. We then analyze black hole temperatures and masses consistent 
with alternative entropies.

\end{abstract}


\maketitle

\section{Introduction} 
\setcounter{equation}{0} 
\label{sec:1}

Einstein said of thermodynamics that ``A theory is the more impressive the 
greater the simplicity of its premises is, the more different kinds of 
things it relates, and the more extended is its area of applicability. 
Therefore the deep impression which classical thermodynamics made upon me. 
It is the only physical theory of universal content concerning which I am 
convinced that within the framework of the applicability of its basic 
concepts, it will never be overthrown'' \cite{Einstein}. Indeed, 
thermodynamics is applied to a large variety of physical theories 
and situations, but its application to systems with long range 
interactions (where the partition function commonly diverges), and to 
gravity in particular, constitutes a challenge. A major discovery in the 
1970s was the formulation of black hole thermodynamics 
\cite{Bardeen:1973gs,Wald:1999vt}. It started with the 
discovery of the Bekenstein-Hawking entropy of black holes 
\cite{Bekenstein:1973ur} $ {\cal S} = \frac{A}{4G}$, where $A$ is the area 
of the event horizon and we use geometrized units in which the speed of 
light $c$, the reduced Planck constant $\hbar$, and the Boltzmann constant 
$K_\mathrm{B}$ are unity. The pieces of the puzzle fell into place when 
Hawking discovered that the Schwarzschild black hole radiates scalar 
quanta at the Hawking temperature 
\be 
T_\mathrm{H} =\frac{1}{8\pi M} \,,\label{HawkingT} 
\ee 
where $M$ is the black hole mass 
\cite{Hawking:1975vcx}. While, in classical thermodynamics, entropy is 
universal and defined uniquely, quantum effects 
correct the Bekenstein-Hawking entropy pointing to its modification in 
full quantum gravity, while inequivalent entropies arise also in 
non-extensive thermodynamics. New entropy proposals come from classical 
gravity as well. Here we summarize recent work and ideas on the 
application of alternative entropies to cosmology and black holes 
\cite{Nojiri:2022aof,Nojiri:2022sfd}.

New entropy proposals in the literature include non-extensive entropies that 
reduce to ${\cal 
S}$ in some limit, such as the Tsallis entropy (\cite{tsallis}, see also 
\cite{Ren:2020djc,Nojiri:2019skr}) for systems with long range 
interactions \begin{align} \label{TS1} \mathcal{S}_\mathrm{T} = 
\frac{A_0}{4G} \left( \frac{A}{A_0} \right)^\delta \,, \end{align} where 
$A_0$ is a constant area and the dimensionless 
parameter $\delta$ quantifies non-extensivity. Other proposals are the 
R{\'e}nyi entropy \cite{renyi, Czinner:2015eyk,Tannukij:2020njz} 
\begin{align} \label{RS1} 
\mathcal{S}_\mathrm{R}=\frac{1}{\alpha} \, \ln 
\left( 1 + \alpha \mathcal{S} \right) \end{align} related to information 
theory, the Sharma-Mittal entropy \cite{SayahianJahromi:2018irq} 
\begin{align} \label{SM} 
\mathcal{S}_\mathrm{SM} = \frac{1}{R}\left[ 
\left(1 + \delta \, \mathcal{S}_\mathrm{T} \right)^{R/\delta} - 1 \right] 
\end{align} 
(with $R$ and $\delta$ free parameters), the Barrow entropy 
\cite{Barrow:2020tzx} 
\begin{align} \label{barrow} 
\mathcal{S}_\mathrm{B} 
= \left( \frac{A}{A_\mathrm{Pl}}\right)^{1+ \, \Delta/2} 
\end{align} 
proposed as a toy model for quantum spacetime foam (where $A_\mathrm{Pl}$ 
is the Planck area), the Kaniadakis entropy 
\cite{Kaniadakis:2005zk} 
\begin{align} \label{kani} 
\mathcal{S}_\mathrm{K} = \frac{1}{K} \sinh \left(K \mathcal{S} \right) 
\end{align} 
generalizing the Boltzmann-Gibbs entropy in relativistic 
statistical systems \cite{Kaniadakis:2005zk}, and the 
non-extensive Loop Quantum Gravity proposal 
\cite{Majhi:2017zao,Czinner:2015eyk} 
\be 
\mathcal{S}_q= \frac{1}{1-q} \left[ \mbox{e}^{ (1-q)\Lambda(\gamma_0) 
{\cal S} } -1 \right] \,,\label{LQGentropy} \ee where the entropic index 
$q$ quantifies how the probability of frequent events is enhanced 
relatively to infrequent ones, $ \Lambda( \gamma_0) = \frac{ \ln 
2}{\sqrt{3} \, \pi \gamma_0} $, and $\gamma_0$ is the Barbero-Immirzi 
parameter, usually taking one of the two values $ \frac{\ln2}{\pi 
\sqrt{3}}$ or $\frac{\ln 3}{2\pi \sqrt{2} }$, depending on the gauge group 
used.
 
These new entropies share four properties, which we promote to minimal 
requirements for any alternative entropy proposal:

\begin{enumerate}

\item {\it Generalized third law:} The entropy vanishes when the 
Bekenstein-Hawking entropy ${\cal S}$ does.  While, in the standard 
thermodynamics of closed systems in equilibrium, $\e^S$ expresses the 
number of states and the entropy $S$ vanishes at zero temperature because 
the ground (vacuum) state should be unique, the Bekenstein-Hawking entropy 
diverges when $T_\mathrm{H} \to 0$ and vanishes as $T_\mathrm{H} 
\rightarrow \infty$.  We require generalized entropies to vanish when the 
Bekenstein-Hawking entropy $\mathcal{S}$ does.

\item {\it Monotonicity:} The entropy is a monotonically increasing 
function of the Bekenstein-Hawking entropy $\mathcal{S}$.

\item {\it Positivity:} The entropy is positive, as the number of states 
$\e^\mathcal{S}$ is greater than unity.

\item {\it Bekenstein-Hawking limit:} The entropy reduces to the 
Bekenstein-Hawking prescription ${\cal S}$ in an appropriate limit.

\end{enumerate}

A new and very general entropy with the above properties and incorporating 
all the above-mentioned entropy proposals as special limits is 
\cite{Nojiri:2022aof} 
\begin{align} \label{general1} 
\mathcal{S}_\mathrm{G} \left( \alpha_\pm, \beta_\pm, \gamma_\pm \right) = 
\frac{1}{\alpha_+ + \alpha_-} \left[ \left( 1 + \frac{\alpha_+}{\beta_+} 
\, \mathcal{S}^{\gamma_+} \right)^{\beta_+} - \left( 1 + 
\frac{\alpha_-}{\beta_-} \, \mathcal{S}^{\gamma_-} \right)^{-\beta_-} 
\right] \,, 
\end{align} 
where we assume all the parameters $\left( \alpha_{\pm} , \beta_{\pm}, 
\gamma_{\pm} \right) $ to be non-negative. 
This proposal reproduces~(\ref{TS1})-(\ref{LQGentropy}) for appropriate 
parameter values \cite{Nojiri:2022aof}. 

A simpler alternative proposal is the 3-parameter entropy 
\cite{Nojiri:2022aof} 
\begin{align} \label{general6} 
\mathcal{S}_\mathrm{G} \left( \alpha, \beta, \gamma \right) = 
\frac{1}{\gamma} \left[ \left( 1 + \frac{\alpha}{\beta} \, \mathcal{S} 
\right)^\beta - 1 \right] \,, 
\end{align} 
where $ \left( \alpha , \beta, 
\gamma \right)$ are non-negative. This quantity reduces to the 
Sharma-Mittal entropy~(\ref{SM}) with $\mathcal{S}_\mathrm{T}=\mathcal{S}$ 
(or $\delta=1$) when $\gamma=\alpha$.  If $\gamma = \left( \alpha/\beta 
\right)^{\beta} $, the limit $\alpha\to \infty$ yields 
\begin{align} \label{general7} 
\lim_{\alpha\to \infty} \mathcal{S}_\mathrm{G} \left( 
\alpha, \beta, \gamma=\left( \frac{\alpha}{\beta} \right)^\beta \right) = 
\mathcal{S}^\beta  
\end{align} 
and the choices $\beta=\delta$ and 
$\beta= 1 + \Delta/2$ reproduce the Tsallis and Barrow 
entropies~(\ref{TS1}) and (\ref{barrow}), respectively. The limit 
$\alpha\rightarrow 0 $, $\beta \rightarrow 0$ with $ \alpha/\beta $ finite 
gives instead the R{\'e}nyi entropy~(\ref{RS1}) (where $\alpha/\beta$ is 
replaced by $\alpha$ and $\gamma=\alpha$). Finally, $\beta\to \infty$ 
and $\gamma=\alpha$ gives the new quantity $ 
\mathcal{S}_\mathrm{G} \left( \alpha, \beta\to \infty, \alpha \right) \to 
\left( \e^{\alpha\mathcal{S}} - 1 \right)/\gamma $ 
satisfying our four entropy requirements.

\section{Black holes and the holographic universe} 
\label{sec:2}

Let us apply the generalized entropy to the Schwarzschild geometry 
\begin{align} 
\label{dS3BB} 
ds^2 = - \left( 1 - \frac{2GM}{r}\right) \, 
dt^2 + \frac{dr^2}{ 1 -2GM/r }
+ r^2 \left( d\vartheta^2 +\sin^2 \vartheta \, d\varphi^2 \right)  \,, 
\end{align} 
where $M$ is the black hole mass.  One can attempt to  identify the 
Tsallis or the R{\' e}nyi entropies~(\ref{TS1}) or~(\ref{RS1}) with the 
black hole entropy \cite{Nojiri:2021czz}; then, if we assume that the mass 
$M$ coincides with the thermodynamical energy $E$ 
\cite{Czinner:2015eyk,Tannukij:2020njz}, 
$d\mathcal{S}_G=dE/T_\mathrm{G}$ requires the temperature $T_\mathrm{G}$ 
to be defined by 
\begin{align} 
\label{TR1} 
\frac{1}{T_\mathrm{G}} \equiv  \frac{d\mathcal{S}_\mathrm{G}}{dM} \neq 
\frac{1}{T_\mathrm{H}} \,.   
\end{align} 
Alternatively, assuming the Hawking temperature~(\ref{HawkingT}) as the 
thermodynamic temperature, the first law 
\be \label{Energy} 
dE_\mathrm{G} = T_\mathrm{H} \, 
  d\mathcal{S}_\mathrm{G} = \frac{\alpha}{\gamma \sqrt{16\pi G}} \left[ 
  \mathcal{S}^{-1/2}
+ \frac{\alpha \left( \beta -1 \right)}{\beta} \, \mathcal{S}^{1/2} + 
  \mathcal{O}\left( \mathcal{S}^{3/2} \right) \right] \ee gives \be 
  \label{GE2} E_\mathrm{G} = \frac{\alpha}{\gamma} \left[ M + \frac{4\pi G 
  \alpha \left( \beta -1 \right)}{3\beta} \, M^3 + \mathcal{O}\left( M^5 
  \right) \right] \neq M \,. 
\ee
The search for new entropies must deal with this problem, which requires a 
deeper re-examination of the broader thermodynamical formalism 
\cite{Nojiri:2022aof}. Black hole thermodynamics is expected to change 
drastically when quantum gravity becomes important. Eventually, the latter 
should change classical and black hole thermodynamics, not only redefining 
entropy but also correcting well-established quantities such as 
temperature and energy.

Thermodynamics has been applied fruitfully to another area of 
gravitational physics, that is, cosmology, to which we now turn our 
attention.  In the holographic dark energy (HDE) scenario 
\cite{Nojiri:2005pu}, thermodynamics plays a primary role since it is 
applied successfully to explain dark energy with the entropy of the 
cosmological horizon. In this 
context, the density of the HDE is proportional to the square of 
the inverse holographic cutoff $L_\mathrm{IR}$, $ 
\rho_\mathrm{hol}=\frac{3C^2}{\kappa^2 {L_\mathrm{IR}}^2} $, where $C$ is 
a free parameter. This holographic cutoff $L_\mathrm{IR}$ is usually taken 
to be the size of the particle horizon $L_\mathrm{p}$ or of the future 
horizon $L_\mathrm{f}$, but there is no compelling argument for choosing 
this quantity. Following \cite{Nojiri:2005pu}, the cutoff is assumed to 
depend on $ L_\mathrm{IR}(L_\mathrm{p}, \dot L_\mathrm{p}, \ddot 
L_\mathrm{p}, \cdots, L_\mathrm{f}, \dot L_\mathrm{f}, \cdots, a)$, which 
gives the ``generalized HDE'' 
\cite{Nojiri:2005pu}. In the spatially flat 
Friedmann-Lema\^{i}tre-Robertson-Walker (FLRW) universe described by the 
line element 
\begin{equation} 
ds^2=-dt^2+a^2(t) \left( dx^2 +dy^2 +dz^2 
\right) \label{metric} 
\end{equation} 
in comoving coordinates, one 
speculates that the generalized HDE originates from the entropy of the 
cosmological horizon. The physical radii of the particle and 
event horizons of the FLRW universe~(\ref{metric})  are $ 
L_\mathrm{p}\equiv a(t) \int_0^t\frac{dt'}{a(t')} $ and $ 
L_\mathrm{f}\equiv a(t) \int_t^\infty \frac{dt'}{a(t')} $ (when these 
integrals converge).  Differentiating gives 
\begin{equation} \label{HLL} 
H \left( L_\mathrm{p} , \dot{L}_\mathrm{p} \right) = 
\frac{\dot{L}_\mathrm{p}}{L_\mathrm{p}} - \frac{1}{L_\mathrm{p}}\, , \quad 
\quad H(L_\mathrm{f} , \dot{L}_\mathrm{f}) = 
\frac{\dot{L}_\mathrm{f}}{L_\mathrm{f}} + \frac{1}{L_\mathrm{f}} \,, 
\end{equation} 
where $H\equiv \dot a/a $, an overdot denoting differentiation with 
respect to $t$.  In the thermodynamical approach 
to gravity ({\em e.g.}, \cite{Padmanabhan:2009vy}), the 
Einstein-Friedmann equations are derived from the Bekenstein-Hawking 
entropy ${\cal S}$:  the apparent horizon of the 
universe~(\ref{metric}) has radius $ r_\mathrm{H}= H^{-1} $, area $A =4\pi 
H^{-2} $, and Bekenstein-Hawking entropy ${\cal S}=\pi H^{-2}/G$. We 
have 
\begin{equation} \label{Tslls2} 
dQ = - dE = -\frac{4\pi}{3} \, 
r_\mathrm{H}^3 \, \dot{\rho} \, dt = -\frac{4\pi}{3H^3} \, \dot{\rho} \, 
dt = \frac{4\pi}{H^2} \left( \rho + P \right) dt 
\end{equation} 
using covariant conservation $ \dot \rho + 3 H \left( \rho + P \right)=0$; 
from the Gibbons-Hawking temperature $T = H/(2\pi)$ and the first law of 
thermodynamics $TdS = dQ$, it follows that $ \dot H = - 4\pi G \left( \rho 
+ P \right) $ and integration gives the Friedmann equation 
\begin{equation} \label{Tslls8} 
H^2 = \frac{8\pi G}{3} \rho + \frac{\Lambda}{3} \,, 
\end{equation} 
where the cosmological constant 
$\Lambda$ appears as an integration constant.  If the Bekenstein-Hawking 
entropy ${\cal S}$ is replaced by a different (non-extensive) concept of 
entropy, the Friedmann equation~(\ref{Tslls8}) is modified, which is 
attributed to the HDE. For example, the Tsallis entropy~(\ref{TS1}) yields 
the modified Friedmann equation 
\begin{align} 
H^2=\frac{8\pi G}{3} \left( 
\rho + \rho_\mathrm{T} \right) + \frac{\Lambda}{3} \,, \quad \quad 
\rho_\mathrm{T} = \frac{3}{8\pi G} \left[ H^2 - \frac{\delta}{2 - 
\delta}\, H_1^2\left( \frac{H}{H_1} \right)^{2(2 - \delta)} \right] \, . 
\label{Tslls11BB} 
\end{align} 
Interpreting $\rho_\mathrm{T} 
=\frac{3C^2}{\kappa^2 {L_\mathrm{IR,T}}^2} $ as the HDE due to the 
infrared holographic cutoff $ L_\mathrm{IR,T}$ leads to  
\begin{eqnarray} 
\label{Tbasic} L_\mathrm{IR,T} = \frac{1}{C} \left[ \left( 
\frac{\dot{L}_\mathrm{f}}{L_\mathrm{f}} + \frac{1}{L_\mathrm{f}} \right)^2
- \frac{\delta}{2 - \delta} \, H_1^2\left( 
\frac{\frac{\dot{L}_\mathrm{f}}{L_\mathrm{f}} + 
\frac{1}{L_\mathrm{f}}}{H_1} \right)^{2(2 -\delta) } \right]^{-1/2} \,. 
\end{eqnarray} 
The Barrow entropy~(\ref{barrow}) describing the spacetime quantum foam 
phenomenologically gives
\begin{align} \label{rhoB} 
\rho_\mathrm{B} = 
\frac{3}{8\pi G} \left[ H^2 - \left( \frac{1 + \Delta/2}{1 - \Delta/2} 
\right)  \frac{16 \pi G}{A_\mathrm{Pl}^2} \left( \frac{H^2}{4\pi 
A_\mathrm{Pl}} \right)^{1 - \Delta/2 } \right] 
\end{align} 
while, with 
the new three-parameter entropy proposal~(\ref{general6}), one obtains 
\begin{align} \label{3G} \rho_\mathrm{G} = \frac{3}{8\pi G} \left[ H^2 - 
\frac{\pi\alpha}{G\beta\gamma\left( 1 - \beta\right)} \left(\frac{G\beta 
H^2}{\pi \alpha} \right)^{2 - \beta} {}_2F_1 \left( 1 - \beta, 2 - \beta, 
3 - \beta;  - \frac{G \beta H^2}{\pi\alpha} \right) \right] \,, 
\end{align}
where ${}_2F_1 \left(a,b, c; z \right) $ is the hypergeometric function.

The formal conservation law $ \dot\rho_\mathrm{G} + 3 H \left( 
\rho_\mathrm{G} + P_\mathrm{G} \right) =0 $ defines the pressure 
$P_\mathrm{G}$ of the HDE and its equation of state parameter 
\begin{align} \label{eos} 
w_\mathrm{G} \equiv &\,  \frac{P_\mathrm{G}}{\rho_\mathrm{G}} = -1 - 
\frac{2\dot H}{3}  \left[ H^2 - \frac{\pi\alpha}{G\beta\gamma\left( 1 - 
\beta\right)} \left(\frac{G\beta H^2}{\pi \alpha} \right)^{2 - \beta} 
{}_2F_1 \left( 1 - \beta, 2 - \beta, 3 - \beta; - \frac{G \beta 
H^2}{\pi\alpha} \right) \right]^{-1} \nonumber\\
& \times \left[ 1 - \frac{2 - \beta}{\gamma\left( 1 - \beta\right)}
\left(\frac{G\beta H^2}{\pi \alpha} \right)^{1 - \beta} {}_2F_1 \left( 1 - 
\beta, 2 - \beta, 3 - \beta; - \frac{G \beta H^2}{\pi\alpha} \right) 
\right. \nonumber \\
& \qquad \left. + \frac{2- \beta}{\gamma\left( 3 - \beta\right)}
\left(\frac{G\beta H^2}{\pi \alpha} \right)^{2 - \beta} {}_2F_1 \left( 2 - 
\beta, 3 - \beta, 4 - \beta; - \frac{G \beta H^2}{\pi\alpha} \right) 
\right] \,. 
\end{align} 
When matter can be neglected and $\Lambda=0$, the 
Friedmann equation $ H^2 = \frac{8\pi G}{3} \rho_\mathrm{G} $ and 
Eq.~(\ref{3G}) yield $ {}_2F_1 =0 $. The zeros $Z_i$ of this 
hypergeometric function are de Sitter universes with Hubble functions 
given by $ Z_i = - \frac{G \beta H^2}{\pi\alpha} $ and effective 
cosmological 
constants $ \Lambda_\mathrm{eff} = \frac{3\pi \alpha Z_i}{G \beta} $. If 
$\Lambda_\mathrm{eff}$ is large, it can cause early universe inflation; if 
it is very small, it may describe the present accelerated expansion; if it 
is slightly larger than the present dark energy, it could potentially 
solve the Hubble tension problem \cite{Riess:2020fzl,Planck:2018vyg}.

Consider the case of a small $Z_1$: the hypergeometric function is 
approximated as 
\begin{eqnarray} 
{}_2F_1 \left( 1 - \beta, 2 - \beta, 3 - 
\beta; - \frac{G \beta H^2}{\pi\alpha} \right)
& \simeq & 1 - \frac{\left( 1 - \beta \right) \left( 2 - \beta \right)}{3-
\beta} \frac{G \beta H^2}{\pi\alpha} \nonumber\\
&&\nonumber\\
&\, & + \frac{\left( 1 - \beta \right) 
\left( 2 - \beta \right)^2}{4 - \beta}\left(\frac{G \beta H^2}{\pi\alpha} 
\right)^2 \,,\label{HGF}
\end{eqnarray} 
then $ Z_1 = - \frac{G \beta H^2}{\pi\alpha} \simeq - 
\frac{\left(3- \beta\right)}{\left( 1 - \beta \right) \left( 2 - \beta 
\right)} $ and $ H^2 \sim \frac{\left(3- \beta\right)\pi\alpha }{\left( 1 
- \beta \right) \left( 2 - \beta \right)G \beta} $. If $ 3 - \beta \sim 
\mathcal{O}\left( 10^{-2n} \right)$, $ \alpha \sim \mathcal{O}\left( 
10^{-2m} \right)$, it is $ H^2 \sim \left( 10^{-n-m + 28}\, \mathrm{eV} 
\right)^2 $ and $n+m= 61$ gives the current dark energy scale $H\sim 
10^{-33}\, \mathrm{eV}$. If another zero $Z_2$ exists with $|Z_2|$ 
slightly smaller than $Z_1$, the effective cosmological constant can 
potentially solve, or alleviate, the Hubble tension problem 
\cite{Riess:2020fzl,Planck:2018vyg}.

In general, the hypergeometric function can have several or infinitely 
many zeros. If there are a root of order unity or a large and negative 
root $Z_i$, then one can obtain the large Hubble rate corresponding to the 
inflationary epoch. Retaining, for illustration, the first three terms in 
Eq.~(\ref{HGF}), 
\begin{align} \label{HGF11} 
\frac{G \beta H^2}{\pi\alpha} 
= & - \frac{4 - \beta}{2\left( 2 - \beta \right)\left( 3- \beta \right)} 
\left( 1 \pm \sqrt{1 - \frac{4 \left( 3 - \beta \right)^2}{\left(4 - 
\beta\right)\left( 1 - \beta \right)} } \, \right) \,,
\end{align} 
and assuming $\beta \lesssim 3$, we have 
\begin{align} 
Z_+ = - \frac{4 - \beta}{2\left( 2 - \beta \right)\left( 3- \beta \right)} 
\, , \quad \quad Z_- = - \frac{3 - \beta}{2\left( 1 - \beta \right)\left( 
2- \beta \right)} \,. 
\end{align} 
For $n+m= 61$ one finds again the present Hubble scale.  
If, instead, $\frac{G \beta H^2}{\pi\alpha} = Z_+$, then $ H^2 \sim \left( 
10^{n-m + 28}\, \mathrm{eV} \right)^2 $ and, for $n+m= 61$, it is $ H^2 
\sim \left( 10^{-2m + 89}\, \mathrm{eV} \right)^2 $. At the GUT scale 
$\sim 10^{16}\, \mathrm{GeV} = 10^{25}\, \mathrm{eV} $ and inflation with 
$H\sim 10^{22}\, \mathrm{eV}$, we obtain $m\sim 33$ or $34$, so $Z_+$ may 
explain early universe inflation.

One can also study generalized HDE from the full six-parameter 
entropy~(\ref{general1}) instead of using the simpler 
proposal~(\ref{general6}), as we did here. Correspondingly, there are many 
more possibilities to realize realistic cosmic histories.

\section{Alternative entropies and corresponding energies}
\label{sec:3}

We describe spherical, static, and asymptotically flat spacetimes with the 
geometry
\begin{align}
\label{metric}
ds^2 = g_{\mu\nu} dx^{\mu} dx^{\nu} 
= -\e^{2\nu(r)} dt^2 + \e^{2\lambda(r)} dr^2
+ r^2 d\Omega_{(2)}^2 \, .
\end{align}
where  $ \lambda(r ) \to 0$  and  $ \nu( r ) \to 0 $ as $r\to +\infty$. 
Let us consider Einstein gravity and interior solutions.   The $t$-$t$  
Einstein equation is
\begin{align}
\label{EinTOV0}
 - \kappa^2 \rho = \frac{1}{r^2} \Big( r \e^{-2\lambda} - r \Big)' \,,
\end{align}
where $\rho$ is the energy density, $f' \equiv df/dr$, and the mass is 
given by 
\begin{equation}
\e^{-2\lambda} \equiv 1 - \frac{\kappa^2 m(r)}{4\pi r} \,,
\end{equation}
\be
4\pi r^2 \rho = m'(r) \,.
\ee
Here 
\begin{align}
\label{TOV1}
m(r) =4 \pi \int^r_{0} r^{\prime 2} \rho (r') dr'  + m_0 \,,
\end{align}
with $m_0$ an integration constant. The metric inside a matter ball  must 
be regular at $r=0$, hence  
\begin{align}
\label{consin}
\lambda\to 0 \, , \quad\quad 
\lambda'(r) =\frac{m-rm'}{r(r-2m)} \xrightarrow{} 0 
\end{align} 
as $r\to0$ or else there is a 
conical singularity. Then, $m_0 =0$ and 
\begin{align}
m(r)=4 \pi \int^ r_{0} r^{\prime 2} \rho (r') dr'  \,.
\label{eq:massfunc}
\end{align}
For asymptotically Schwarzschild geometries 
\begin{align}
\label{mass}
M=m(r\to\infty)=4 \pi \int_0^\infty d r \, r^2 \rho(r) 
\end{align}
while, in the presence of a central singularity, the integration 
constant $m_0$ remains and
\begin{align}
\label{massB}
M=m(r=\infty)=4 \pi \int_0^\infty d r\, r^2 \rho(r) + m_0 \,.
\end{align}
The total mass is not  $m(r=\infty)$ but  \cite{Nojiri:2022sfd}
\begin{eqnarray}
\bar{M} &=& \int d^3 x \, \sqrt{\gamma} \, \rho(r) \\
&=& 4 \pi \int_0^\infty \rho(r) r^2 \left[1-\frac{2 G m(r)} 
r\right]^{-1/2}  d r \nonumber\\
& = & 4 \pi \int_0^{\infty} dr \, \rho(r) r^2 \left[1 + \frac{G m(r)} r - 
\frac{3G^2 m^2(r)}{r^2} + \mathcal{O} \left( G^3 \right) \right] 
\,, \label{barM}
\end{eqnarray} 
where $\gamma$ is the determinant of the 3D Riemannian metric 
\begin{align}
\label{induced3metric}
\gamma_{\ell m} \, d x^{\ell} d x^m =\e^{2 \lambda} d r^{2}+r^{2} 
d\Omega_{(2)}^2 \, .
\end{align}
The gravitational binding energy of the ball is $ E_{\mathrm{B}}=M-\bar{M} $.
The second  term in the last line of Eq.~(\ref{barM}) is interpreted as the  
Newtonian gravitational potential energy 
\begin{align}
\label{NP}
 - 4 \pi G \int_0^\infty dr \, \rho(r) \, r^2 \frac{m(r)} r = - 
\frac{G}{2} \int 
dV \int dV' \, \frac{\rho\left(\bm r\right) \rho\left(\bm 
r'\right)}{\left| 
\bm r - \bm r' \right|} \, ,
\end{align}
where  the general-relativistic corrections are of order $G^2$ and 
higher.

If a black hole geometry is asymptotically Schwarzschild we can impose $ 
m(r\to \infty)=M$, fixing the integration constant $m_0$ (one obtains 
$m_0=M$ for the Schwarzschild black hole, for which $\rho$ can be seen as 
proportional to a Dirac delta centered at $r=0$). The mass $M$ coincides 
with the Arnowitt-Deser-Misner mass.

Let us consider now modified gravity, in which case we write the $t$-$t$ 
field equation as 
\begin{align}
\label{MGTOV}
 - \kappa^2 \rho_\mathrm{eff} = \frac{1}{r^2} \left( r \e^{-2\lambda} - r 
\right)' \,.
\end{align}
Now the effective energy density $ \rho_\mathrm{eff}$ is defined by 
casting the field equations as effective Einstein equations with 
right-hand sides that contain  effective stress-energy tensors built 
with the non-Einsteinian terms. Now the effective mass is 
\begin{align}
\label{MDmassfunc}
m_\mathrm{eff}(r)=4 \pi \int^r_0dr' r^{\prime 2} \rho_\mathrm{eff} 
(r')\,.
\end{align}

For example, for $F(R)$ gravity  
\begin{align}
\label{FR}
S_{F(R)} =\frac{1}{2\kappa^2} \int d^4 x \sqrt{-g} \, F(R) + 
S^\mathrm{(matter)}
\end{align}
(where $R$ is the Ricci scalar,  $F(R)$ is a nonlinear function,  
and $g$ is the determinant of the metric $g_{\mu\nu}$), we write 
$F(R) \equiv R+f(R)$ and $f_R(R) \equiv df(R)/dR$. The $\left(0,0 \right)$ 
field equation  defines the total energy density $ 
\rho_\mathrm{eff} = \rho + \rho_{F(R)} $, where 
\begin{align}
\rho_{F(R)} \equiv &\, \frac{1}{\kappa^2} \left\{ 
 - \frac{f}{2} - \e^{- 2 \lambda} \left[
\nu'' + \left(\nu' - \lambda'\right)\nu' + \frac{2\nu'} r\right] f_R 
\right.
\nonumber\\
&\left. 
+ \e^{ -2\lambda} \left[ f_R'' + \left( - \lambda' + \frac{2} r \right) 
f_R' \right] \right\} \,.
\end{align}
The resulting (effective) total mass  
\begin{align}
\label{barMeff}
\bar{M}_\mathrm{eff} = \int d^3 x \, \sqrt{\gamma}\, \rho_\mathrm{eff}(r) 
= \int d^3 x \, \sqrt{\gamma} \left( \rho + \rho_{F(R)} \right) \, . 
\end{align}
receives contributions from both matter and gravity.  The leading correction 
to the binding energy is 
\begin{align}
\label{bindingeff}
E_{\mathrm{B,eff}} = - G \int dV \int dV' \, \frac{\left[ \rho\left(\bm 
r\right) + \rho_{F(R)}\left(\bm r\right) \right]
\left[ \rho\left(\bm r'\right) + \rho_{F(R)}\left(\bm r'\right) 
\right]}{ \left| \bm r - \bm r' \right|} + \cdots \,,
\end{align}
$M_\mathrm{eff}\equiv m_\mathrm{eff}\left( 
r\to\infty \right)$ is the total mass-energy of the system, while 
$m_\mathrm{eff}(r)$ is the mass-energy of a 2-sphere of radius 
$r$.

A black hole in alternative theories of gravity may have horizon 
radius $r_\mathrm{h} \neq 2G M_\mathrm{eff}  \equiv 
2 m_\mathrm{eff} \left(r\to 
\infty\right) $. Now, if $M_\mathrm{eff}$ is used as the 
internal energy and  $\mathcal{S}=4\pi {r_\mathrm{h}}^2 /4$ as the black 
hole entropy, the new temperature given by  
\begin{align}
\label{temperature}
\frac{1}{T}=\frac{d\mathcal{S}}{dM_\mathrm{eff}}
\end{align}
differs from the usual Hawking temperature $T_\mathrm{H}$. 
Alternatively, if the Hawking temperature is used, the entropy 
\begin{align}
\label{entropy}
d\mathcal{S} = \frac{dM_\mathrm{eff}}{T_\mathrm{H}}
\end{align}
must replace the Bekenstein-Hawking entropy. The difference 
$M_\mathrm{eff} -  m_\mathrm{eff}\left( r_\mathrm{h} \right)$ could then 
be identified with the energy 
outside the horizon. For  this black 
hole, $m_\mathrm{eff}\left( r_\mathrm{h} 
\right)$ would be the internal energy and Eq.~(\ref{entropy}) would become   
\begin{align}
\label{entropy2}
d\mathcal{S}_\mathrm{bh} = \frac{d m_\mathrm{eff}\left( r_\mathrm{h} 
\right)}{T_\mathrm{H}}\,.
\end{align}

\section{Temperatures corresponding to alternative entropies}
\label{sec:4}

Denote the metric coefficients as 
\begin{align}
\label{ss1}
h(r) \equiv \e^{2\nu(r)} \, , \quad \quad h_1(r) \equiv \e^{-2\lambda(r)} 
\,,
\end{align}
then the roots of $h(r)=0$ locate the event horizon. If $h_1(r)$ does not 
vanish simultaneously with $h(r)$, the spacetime curvature diverges as  
$h(r) \to 0$. If $h_1(r)$ and $h(r)$ do vanish simultaneously, the surface 
$ 
h_1(r)=h(r) =0$ is an event  horizon. In fact, consider 
the curvature invariants 
\begin{align}
\label{inv33}
R_{\mu \nu \rho \sigma} R^{\mu \nu \rho \sigma} =&\, \frac {1}{4h^4 
r^4}\left[ 4 r^4 h''^2 h^2 {h_1}^2
+4 r^4hh_1 h' h'' \left( h'_1 h - h' h_1 \right)  +  \left( h'^2 {h_1} 
r^2\right)^2  \right. \nonumber\\
& \left. -2 r^4 h'^3h_1h'_1h  + \left(r h {h'}\right)^2 \left( {h_1'}^2 
r^2+8  {h_1} \right)^2   \right.\nonumber\\
& \left. +8 \, h^4 
\left(  r^2 {h'_1}^2 + 2\, \left( 1- h_1 \right)^2 \right) \right] \,, \\
&\nonumber\\
R_{\mu \nu } R^{\mu \nu }=&\, \frac {1}{8h^4 r^4}\left[ 4 r^4 h''^2h^2 
{h_1}^2+4 h 
\left[h \left( rh'_1 +2h_1 \right)h'- r h'^2h_1 +2\ h^2h'_1 \right] 
r^3h_1 h'' \right. \nonumber\\
& + r^4h'^4{h_1}^2  + r^2h^2 \left( 12 {h_1}^2+{h'_1}^2 r^2 \right)h'^2 -2 
r^3h h_1 \left( 
rh'_1 +2 h_1 \right) h'^3 \nonumber\\
& \left. +4r h^3 \left( 2h'_1rh_1 -4h_1 +4 {h_1}^2+ {h'_1}^2 r^2 \right) 
h' \right. \nonumber\\
& \left. 
+4h^4 \left(3 {h'_1}^2 r^2 +4r \left( h_1-1 \right)h'_1 +4\left(h_1 
-1 \right)^2 \right) \right] \,,\\
&\nonumber\\
R=&\, \frac{2 h'' h_1 h r^2- r^2h_1  h'^2 +r h' h \left( rh'_1+ 4h_1   
\right) +4h^2 \left( h_1 + rh'_1  -1 \right) }{2 h^2 r^2 }\,.
\end{align}
Their denominators contain positive 
powers of $h$ and these invariants diverge as $h \to 
0$. If $h_1(r)$ and $h(r)$ vanish simultaneously, 
the  invariants~(\ref{inv33}) remain finite where $ h_1(r) = h(r) =0$  
since  $h_1(r) = h_2(r) h(r) $ and   $h_2 \neq 0 $ and is regular 
where $h(r)=0$. Then the substitution of  $h_1 = h_2 h$ in 
Eqs.~(\ref{inv33}) yields 
\begin{align}
\label{invariants2}
R_{\mu \nu \rho \sigma} R^{\mu \nu \rho \sigma}
=&\, {h''}^2 {h_2}^2 + h' h_2 h_2'  + \left(  \frac{ {h'} {h_2'} }{2} 
\right)^2  + 
\left( \frac{2 {h_2} {h'}}{r} \right)^2 \nonumber\\
& + \frac{2\left[ r^2 \left( h h_2'  + h' h_2  \right)^2  
+ 2 \left( {h_2}^2 h^2 - 1 \right)^2 \right] }{r^4} \, , \\
&\nonumber\\
\label{invariants3}
R_{\mu \nu } R^{\mu \nu } =&\, \frac{{h''}^2 {h_2}^2}{2} + \frac{h'' h' 
h_2 h_2'}{2} + \frac{{h_2}^2 h' h''  } r + \frac{ h_2 h^2  \left( h h_2'  
+  h_2 h'\right) }{r} 
+ \frac{3 {h_2}^2 {h'}^2 }{2r^2} \nonumber\\
& + \frac{{h'}^2 {h_2}^2 }{8} + \frac{ h_2 h'  
\left( h h_2' + h_2 h' \right)}{r^2} \nn
& - \frac{2 {h_2}^2 h' }{r^3} + \frac{2h {h_2}^2 h' }{r^3} + \frac{h  
{h_2}^2 h' }{2r}  + \frac{ h_2  {h'}^2 h_2'} r \,\nonumber\\
& \times \, 
 \frac{3 \left( h h_2' + h_2 h' \right) r^2
+4r \left( h h_2 -1 \right) \left( h h_2' + h_2 h' \right) +4\left(1-h h_2 
 \right)^2}{2r^4} \, , 
\end{align}
and
\begin{align}
\label{invariants1}
R=&2\, h_2  h'' + \frac{2 \, r h_2 h'}{r} + \frac{ h' h_2' }{2} + 
\frac{2\left[ h_1+   r \left( h h_2'  +  h_2 h'\right)  - 1\right]}{r^2} 
\,,
\end{align}
and these invariants remain finite as $h(r) \to 0$.

Given that $h_1(r)$ and $h(r)$ vanish simultaneously on the event horizon, 
we can write  $h_1(r) =  \e^{-2\lambda(r)}$ and the radius of the event 
horizon is  
\begin{align}
\label{horizon}
r_\mathrm{h}= \frac{\kappa^2 m(r_\mathrm{h})}{4\pi}= 2G m(r_\mathrm{h}) 
\,.
\end{align}
Close to the horizon, {\em i.e.}, at $r \equiv r_\mathrm{h} + \delta r$,  
\begin{eqnarray}
\e^{-2\lambda} &=& h_1 = h h_2  = \frac{C\left( r_\mathrm{h} 
\right) \left( r - r_\mathrm{h} \right)   }{r_\mathrm{h}} \,,\\ 
\e^{2\nu} &=& h = \frac{h_1}{h_2} = \frac{C\left( r_\mathrm{h} 
\right) \left( r - r_\mathrm{h} \right)  }{h_2 \left( r_\mathrm{h} \right) 
r_\mathrm{h}}  \,,
\end{eqnarray} 
where $C\left( r_\mathrm{h} \right)\equiv 1 - m'\left( r_\mathrm{h} 
\right)$. Wick-rotating the time  $t\to i\tau$, the 
near-horizon geometry~(\ref{metric}) becomes
 \begin{align}
\label{TH1}
ds^2 \simeq \frac{C\left( r_\mathrm{h} \right) \delta r}{h_2 \left( 
r_\mathrm{h} \right) r_\mathrm{h}} \, d\tau^2 
+ \frac{r_\mathrm{h}}{C\left( r_\mathrm{h} \right) \delta r} \, d(\delta 
r)^2 
+ r_\mathrm{h}^2 \, d\Omega_{(2)}^2 \,.
\end{align}
Introduce the new radial coordinate  
defined by $ d\rho = d \left( \delta r\right) \sqrt{ 
\frac{r_\mathrm{h}}{C\left( r_\mathrm{h} 
\right) \delta r}} $ and   
\begin{align}
\label{TH2}
\rho = 2 \, \sqrt{ \frac{r_\mathrm{h} \, \delta r}{C\left( r_\mathrm{h} 
\right)}} \quad\quad \delta r= \frac{C\left( r_\mathrm{h} \right) 
\rho^2}{4 r_\mathrm{h}} \,,
\end{align}
then the near-horizon geometry~(\ref{TH1}) reads
\begin{align}
\label{TH3}
ds^2 \simeq \frac{C\left( r_\mathrm{h} \right)^2}{ 4 h_2 \left( 
r_\mathrm{h} \right) r_\mathrm{h}^2} \, \rho^2d\tau^2
+ d\rho^2 + r_\mathrm{h}^2 \, d\Omega_{(2)}^2 \,.
\end{align}
To avoid conical singularities near $\rho = 0$, one imposes that the 
Euclidean time  $\tau$ is periodic of period $t_*$,
\begin{align}
\label{TH4}
\frac{C\left( r_\mathrm{h} \right) \tau}{ 2 r_\mathrm{h} \sqrt{h_2 \left( 
r_\mathrm{h} \right)}} 
\simeq \frac{C\left( r_\mathrm{h} \right) \tau}{ 2 \, r_\mathrm{h} \, 
\sqrt{h_2 \left( r_\mathrm{h} \right)}} + 2 \, \pi \,.
\end{align}
As a result, the temperature corresponds to $t_*^{-1}$. In the 
Euclidean path integral formulation of finite-temperature field theory 
\begin{align}
\label{TH5}
\int \left[ D\phi \right] \e^{ \int_0^{t_*}  L(\phi)} dt 
= \, \mathrm{Tr}\left( \e^{-t_* H} \right) = \, \mathrm{Tr} \left( \e^{ - 
\frac{H}{T} } \right) 
\end{align}
and the temperature of the Schwarzschild black hole  
\begin{align}
\label{TH6}
T = \frac{C\left( r_\mathrm{h} \right)}{4\pi r_\mathrm{h}\sqrt{h_2 \left( 
r_\mathrm{h} \right)}} 
= \frac{C\left( r_\mathrm{h} \right)}{8\pi G 
m_\mathrm{eff} \left( r_\mathrm{h} \right) \sqrt{h_2 \left( r_\mathrm{h} 
\right)} } 
= \frac{C\left( r_\mathrm{h} \right) T_\mathrm{H}}{\sqrt{h_2 \left( 
r_\mathrm{h} \right)}} 
\end{align}
follows which, in general,  differs from the Hawking temperature  
\begin{equation}
T_\mathrm{H}\equiv \frac{1}{8\pi G m_\mathrm{eff} \left( r_\mathrm{h} 
\right)} 
\end{equation}
by the factor 
$\frac{C \left( r_\mathrm{h} \right)}{\sqrt{h_2 
\left( r_\mathrm{h} \right)}}$ that cannot be absorbed into a  
time rescaling  because we have fixed the scale so that 
\begin{equation}
h\left( r \to \infty 
\right) = h_2\left( r \to \infty \right) h_1\left( r \to \infty \right) 
=\e^{2\nu\left( r \to \infty \right) } =1 \,.
\end{equation}
Since Hawking radiation is a near-horizon phenomenon, 
thermal radiation can correspond to the  temperature~(\ref{TH6}). 

Identifying $m_\mathrm{eff} 
\left( r_\mathrm{h} \right)$ with the black hole internal energy, 
Eq.~(\ref{entropy2}) yields 
\begin{align}
\label{entropy3B}
\mathcal{S}_\mathrm{bh} = \int \frac{d m_\mathrm{eff} \left( r_\mathrm{h} 
\right)}{T}\, .
\end{align}
The solutions of the gravitational field equations contain integration 
constants $c_i$ for $ i=1, \, \cdots \, , N $ (for example, in general 
relativity (GR) the 
mass  $M$ of the Schwarzschild black hole appears as an integration 
constant in the metric 
coefficients $ \e^{2\nu}=\e^{-2\lambda}= 1 - 2M/r $ when integrating the 
Einstein equations for spherical and asympotically flat vacuum).
$N$ depends on the theory and  $\lambda(r)$, $\nu(r)$,  $m(r)$, $h(r)$, 
and $h_{1,2}(r)$ depend on the integration constants $c_i$.  
The solution $ r_\mathrm{h} 
\left(c_i\right)$  of Eq.~(\ref{horizon}) also depends on  
these integration constants (again, for the Schwarzschild black hole of 
GR, $r_\mathrm{h}=2M$). 
Equation~(\ref{horizon}) yields 
\be
m\left(r_\mathrm{h} \right) = m\Big(  r=r_\mathrm{h} \left(c_i\right); 
c_i\Big)=\frac{r_\mathrm{h} \left(c_i\right)}{2G} \,,
\ee
then the integration constants $c_i$'s can be parametrized with a single 
parameter $\xi$, $c_i=c_i(\xi)$ (for example, the Reissner-Nordstr\"{o}m 
black hole can be parametrized by the charge-to-mass ratio).

In this way, Eq.~(\ref{TH6}) turns  Eq.~(\ref{entropy3B}) into 
\begin{align}
\label{entropy3c}
\mathcal{S}_\mathrm{bh} = \frac{1}{2G} \int d\xi\, \frac{\left[ 4\pi 
r_\mathrm{h} \left( c_i \left(\xi\right) \right) 
\sqrt{ h_2\left(r=r_\mathrm{h} \left(c_i\left(\xi\right) \right); 
c_i\left(\xi\right)\right) } \right]}
{1 - \left. \frac{\partial m\left(r; c_i\left(\xi\right) \right)}{\partial 
r}\right|_{r= r_\mathrm{h} \left( c_i \left(\xi\right) \right)}}
\sum_{i=1}^N \frac{\partial r_\mathrm{h} \left(c_i\right)}{\partial c_i} 
\, \frac{\partial c_i}{\partial \xi}
\end{align}
and choosing  $\xi=r_\mathrm{h}$, Eq.~(\ref{entropy3c}) becomes
\begin{align}
\label{entropy3d}
\mathcal{S}_\mathrm{bh} = \frac{1}{2G} \int_0^{r_\mathrm{h}} d\xi \, 
\frac{\left(4\pi \xi \sqrt{ h_2\left(r=\xi; c_i\left(\xi\right)\right) } 
\right) }{1 - \left. \frac{\partial m\left(r; c_i\left(\xi\right) 
\right)}{\partial r}\right|_{r= \xi}} \,,
\end{align}
where the integration constant is determined by the condition 
$\mathcal{S}_\mathrm{bh} \left( r_\mathrm{h} =0\right) =0$. In GR, the 
Schwarzschild black hole with $h_2(x)=1$, $m=M=$~const. is characterized 
by  the 
Bekenstein-Hawking entropy. The different choice in 
which $h_2\left( r 
\to  r_\mathrm{h} \right)$ gives a  contribution leads to an entropy 
$\mathcal{S}_\mathrm{bh}$ potentially different  from the 
Bekenstein-Hawking one.

According to Eq.~(\ref{entropy3d}),  
\begin{align}
\label{entropy5}
\frac{h_2\left(r=r_\mathrm{h}; c_i\left(r_\mathrm{h}\right)\right) } 
{\left(1 - \left. \frac{\partial m\left(r; c_i\left(r_\mathrm{h}\right) 
\right)}{\partial r}\right|_{r= r_\mathrm{h}}\right)^2}
= 16G^2 \Big[ \mathcal{S}_\mathrm{bh}' \left(A\right) \Big]^2
\end{align}
and various entropy choices  lead to  corresponding forms of 
\begin{equation}
\frac{h_2\left(r=r_\mathrm{h}; 
c_i\left(r_\mathrm{h}\right)\right) } 
{\left(1 - \left. \frac{\partial m\left(r; c_i\left(r_\mathrm{h}\right) 
\right)}{\partial r}\right|_{r= r_\mathrm{h}}\right)^2} \,.
\end{equation}

In \cite{Nojiri:2022aof}, we proposed two generalizations of 
entropy.  We begin with the six-parameter entropy 
\begin{align}
\label{general1}
\mathcal{S}_\mathrm{G} \left( \alpha_\pm, \beta_\pm, \gamma_\pm \right)
= \frac{1}{\alpha_+ + \alpha_-}
\left[ \left( 1 + \frac{\alpha_+}{\beta_+} \, \mathcal{S}^{\gamma_+} 
\right)^{\beta_+} - \left( 1 + \frac{\alpha_-}{\beta_-} 
\, \mathcal{S}^{\gamma_-} \right)^{-\beta_-} \right] \,,
\end{align}
where we take all the parameters $\left( \alpha_{\pm} , 
\beta_{\pm}, \gamma_{\pm} \right) $ to be positive. Adjusting these 
parameters to suitable values, this entropy function  
reduces to the entropies~(\ref{TS1}), (\ref{RS1}), (\ref{SM}), 
(\ref{barrow}), (\ref{kani}), and (\ref{LQGentropy}). 
If we choose  $\alpha_+=\alpha_- = 0$ and  
$\gamma_{-}= \gamma_{+} \equiv \gamma$, the  values 
$\gamma=\delta$ or $\gamma= 1 + \Delta/2 $ give back the Tsallis 
entropy~(\ref{TS1}) and the Barrow entropy (\ref{barrow}).  If 
we pick $\alpha_-=0$ and we write  $\alpha_+=R$, $\beta_+ = R/\delta$, and 
$\gamma_+=\delta$, then we recover the Sharma-Mittal entropy~(\ref{SM}). 
Another possibility consists of the limit $\alpha_+ 
\rightarrow 0$ and $\beta_+ \rightarrow 0$ with $\alpha \equiv \alpha_+ / 
\beta_+ $ 
finite. Further setting $\gamma_+=1$, this procedure  recovers the 
R{\'e}nyi 
entropy~(\ref{RS1}). In the different  limit $\beta_\pm\to 0$ of the 
entropy~(\ref{general1}) with $\gamma_\pm=1$ and $\alpha_\pm = K$ 
the latter is reduced to the Kaniadakis entropy~(\ref{kani}). 
Finally,  if we fix $\alpha$ and $\gamma$ to the values $\alpha_{-}=0 $ 
and  $\gamma_{+}=1$ in Eq.~(\ref{general1}), the limit 
$\beta_{+} \rightarrow +\infty $ in conjunction with $\alpha=1-q$ 
reproduces the Loop Quantum Gravity entropy~(\ref{LQGentropy}) with 
$\Lambda( \gamma_0 ) =1$. 


Another proposal in \cite{Nojiri:2022aof},  containing only 
three parameters, consists of  
\begin{align}
\label{general6}
\mathcal{S}_\mathrm{G} \left( \alpha, \beta, \gamma \right)
= \gamma^{-1} \left[ \left( \frac{\alpha}{\beta} \, \mathcal{S} +1  
\right)^\beta - 1 \right] \,.
\end{align}
We choose again positive values of the parameters $ \alpha , \beta$, and $ 
\gamma $. When 
$\gamma=\alpha$, $\mathcal{S}_\mathrm{G}$ is the same as the 
Sharma-Mittal entropy~(\ref{SM}) with 
$\mathcal{S}_\mathrm{T}=\mathcal{S}$ and $\delta=1$. If we fix $\gamma = 
\left( \alpha/\beta \right)^{\beta} $, then~(\ref{general6}) 
becomes the Tsallis entropy proposal~(\ref{TS1}) if $\beta=\delta$ and the 
Barrow entropy~(\ref{barrow}) if $\alpha\to \infty$. To conclude, the 
limit $\left( \alpha, \beta \right) \rightarrow \left( 0, 0 \right) $  
with $ 
\alpha/\beta $ 
finite yields the R{\'e}nyi entropy~(\ref{RS1}), provided that 
we substitute $\alpha$ in place of  $\alpha/\beta$ and that 
$\gamma=\alpha$.

Let us come to discuss spherical spacetimes while using the 
Tsallis entropy (\ref{TS1}). Equation~(\ref{entropy5}) then becomes
\begin{align}
\label{entropy6}
\frac{h_2\left(r=r_\mathrm{h}; c_i\left(r_\mathrm{h}\right)\right) } 
{\left(1 - \left. \frac{\partial m\left(r; c_i\left(r_\mathrm{h}\right) 
\right)}{\partial r}\right|_{r= r_\mathrm{h}}\right)^2}
= \delta^2 \left( \frac{4\pi {r_\mathrm{h}}^2}{A_0} 
\right)^{2\left(\delta - 1 \right)} \,.
\end{align}
In the same geometry, the R{\'e}nyi entropy construct~(\ref{RS1}) yields 
instead  
\begin{align}
\label{entropy8}
\frac{h_2\left(r=r_\mathrm{h}; c_i\left(r_\mathrm{h}\right)\right) } 
{\left(1 - \left. \frac{\partial m\left(r; c_i\left(r_\mathrm{h}\right) 
\right)}{\partial r}\right|_{r= r_\mathrm{h}}\right)^2}
= \frac{1}{ \left( 1 + \frac{\pi \alpha {r_\mathrm{h}}^2}{G} \right)^2} 
\, .
\end{align}
By contrast, the Kaniadakis entropy~(\ref{kani}) yields 
\begin{align}
\label{kani2}
\frac{h_2\left(r=r_\mathrm{h}; c_i\left(r_\mathrm{h}\right)\right) } 
{\left(1 - \left. \frac{\partial m\left(r; c_i\left(r_\mathrm{h}\right) 
\right)}{\partial r}\right|_{r= r_\mathrm{h}}\right)^2}
= \cosh^2 \left(\frac{\pi K {r_\mathrm{h}}^2}{G} \right) \,,
\end{align}
while our six-parameter entropy~(\ref{general1}) produces 
\begin{align}
\label{general1h2}
\frac{h_2\left(r=r_\mathrm{h}; c_i\left(r_\mathrm{h}\right)\right) } 
{\left(1 - \left. \frac{\partial m\left(r; c_i\left(r_\mathrm{h}\right) 
\right)}{\partial r}\right|_{r= r_\mathrm{h}}\right)^2} 
=\, & \frac{1}{\left(\alpha_+ + \alpha_-\right)^2}
\left[ \alpha_+ \gamma_+ \left(\frac{\pi {r_\mathrm{h}}^2}{G} 
\right)^{\gamma_+ -1}\left( 1 + \frac{\alpha_+}{\beta_+} 
\left(\frac{\pi {r_\mathrm{h}}^2}{G} \right)^{\gamma_+} \right)^{\beta_+ 
- 1} \right. \nonumber \\
&\, \left. + \alpha_- \gamma_- \left(\frac{\pi {r_\mathrm{h}}^2}{G} 
\right)^{\gamma_- -1}\left( 1 + \frac{\alpha_-}{\beta_-} 
\left(\frac{\pi {r_\mathrm{h}}^2}{G} \right)^{\gamma_-} \right)^{-\beta_- 
- 1} \right]^2 \,.
\end{align}
We can also consider our simplified three-parameter 
entropy~(\ref{general6}), which gives  
\begin{align}
\label{general5h2}
\frac{h_2\left(r=r_\mathrm{h}; c_i\left(r_\mathrm{h}\right)\right) } 
{\left(1 - \left. \frac{\partial m\left(r; c_i\left(r_\mathrm{h}\right) 
\right)}{\partial r}\right|_{r= r_\mathrm{h}}\right)^2}
= \frac{\alpha^2}{\gamma^2} \left[ 1 + \left(\frac{\pi\alpha 
{r_\mathrm{h}}^2}{\beta G} \right) \right]^{2\beta - 2} \,.
\end{align}

Specific models realizing these relations have been discussed in 
\cite{Nojiri:2022sfd}.

\section{Conclusions} 
\label{sec:5}

The Bekenstein-Hawking entropy is modified by quantum gravity 
phenomenology, as exemplified by the Barrow and the Loop Quantum Gravity 
proposals~(\ref{barrow}) and~(\ref{LQGentropy}), or by non-extensive 
thermodynamics. While specific modifications abound in the literature and 
may be questionable, the general idea of departures from the simpler 
Bekenstein-Hawking prescription in the presence of phenomena such as 
spacetime foam, loops, or the unusual weighting of frequent/infrequent 
states, appears reasonable. Lacking knowledge of the ``correct'' entropy, 
we propose a phenomenological prescription which incorporates many recent 
and older entropies proposed in the literature and embodies four key 
properties that we identify as essential requirements for any physically 
reasonable entropy. Our most general construct contains six parameters, 
but a simplified version limited to three parameters seems to achieve the 
same goals, as shown in \cite{Nojiri:2021czz,Nojiri:2022aof,Nojiri:2022sfd} 
and summarized here.

In addition to containing the previous Barrow, Loop Quantum Gravity, 
R{\'{e}}nyi, Tsallis, Sharma-Mittal, and Kaniadakis entropies as special 
cases, and to reducing to the Bekenstein-Hawking entropy in an appropriate 
limit, our new proposal exhibits interesting phenomenology when applied to 
holographic dark energy in cosmology. In this context, there is the 
possibility of generating an effective cosmological constant, which can 
either cause early universe inflation or alleviate the current Hubble 
tension afflicting the standard $\Lambda$CDM cosmological model 
\cite{Riess:2020fzl,Planck:2018vyg}. Even though tiny, Planck-scale 
suppressed, infrared corrections to low-energy physics could at first 
sight seem unable to generate observable cosmological effects, this may 
not be the case. While the details of possible quantum 
gravity corrections remain obscure, one can follow Einstein's insight on 
the wide applicability of thermodynamics in physics and search for these 
corrections through their effects on entropy and thermodynamics. The new 
entropy proposals outlined here and in \cite{Nojiri:2022aof} seem to offer 
a practical implementation of this approach to cosmology and gravity.

Changing the notion of entropy jeopardizes the thermodynamics, unless the 
temperature and mass ({\em i.e.}, internal energy) are also changed in a 
suitable way. We have proposed ways of making the entire thermodynamics 
consistent with alternative entropies, but we have not exhausted all 
possibilities. Alternatives will be explored in the future.

\section*{Acknowledgments}

This work is partially supported by JSPS Grant-in-Aid for Scientific 
Research (C) No. 18K03615 (S.~N.), by MINECO (Spain) project 
PID2019-104397GB-I00 (S.~D.~O),  and by the Natural Sciences and 
Engineering Research Council of Canada grant~2016-03803 to V.~F..



\begin{thebibliography}{99}


\bibitem{Einstein} A. Einstein, {\em Autobiographical Notes}, translated 
and edited by P.A. Schilpp, Open Court, La~Salle \& Chicago, Illinois, 
USA (1979).

\bibitem{Bardeen:1973gs} J.M.~Bardeen, B.~Carter and S.W.~Hawking,
Commun. Math. Phys. \textbf{31}, 161 (1973).

\bibitem{Wald:1999vt} R.M.~Wald,
Living Rev. Rel. \textbf{4}, 6 (2001).


\bibitem{Bekenstein:1973ur} J.D.~Bekenstein,
Phys. Rev. D \textbf{7}, 2333 (1973).

\bibitem{Hawking:1975vcx} S.W.~Hawking,
Commun. Math. Phys. \textbf{43}, 199 (1975) [erratum: 
\textbf{46}, 206 (1976)].

\bibitem{Nojiri:2022aof}
S.~Nojiri, S.~D.~Odintsov and V.~Faraoni,
Phys. Rev. D \textbf{105}, no.4, 044042 (2022).

\bibitem{Nojiri:2022sfd}
S.~Nojiri, S.~D.~Odintsov and V.~Faraoni,
[arXiv:2207.07905 [gr-qc]].


\bibitem{tsallis} C.~Tsallis,
J. Stat. Phys. 52, 479-487 (1988).

\bibitem{Ren:2020djc} J.~Ren,
JHEP \textbf{05}, 080 (2021).

\bibitem{Nojiri:2019skr} S.~Nojiri, S.D.~Odintsov and E.N.~Saridakis,
Eur. Phys. J. C \textbf{79}, 242 (2019).

\bibitem{renyi} A.~R{\'{e}}nyi,
Proceedings of the Fourth Berkeley Symposium on Mathematical Statistics 
and Probability, J. Neyman ed., University of California Press (1961), 
547-561.

\bibitem{Czinner:2015eyk} V.G.~Czinner and H.~Iguchi,
Phys. Lett. B \textbf{752}, 306 (2016).

\bibitem{Tannukij:2020njz} L.~Tannukij {\em et al.}, 
Eur. Phys. J. Plus \textbf{135}, 500 (2020).



\bibitem{SayahianJahromi:2018irq} A.~Sayahian Jahromi {\em et al.},
Phys. Lett. B \textbf{780}, 21 (2018).

\bibitem{Barrow:2020tzx} J.D. Barrow,
Phys. Lett. B \textbf{808}, 135643 (2020).

\bibitem{Kaniadakis:2005zk} G.~Kaniadakis,
Phys. Rev. E \textbf{72}, 036108 (2005).


\bibitem{Majhi:2017zao} A.~Majhi,
Phys. Lett. B \textbf{775}, 32 (2017).

\bibitem{Nojiri:2021czz} S.~Nojiri, S.D.~Odintsov and V.~Faraoni,
Phys. Rev. D \textbf{104}, 084030 (2021).

\bibitem{Nojiri:2005pu} S.~Nojiri and S.D.~Odintsov,
Gen.\ Rel.\ Grav.\ {\bf 38}, 1285 (2006).


\bibitem{Padmanabhan:2009vy} T.~Padmanabhan,
Rept. Prog. Phys. \textbf{73}, 046901 (2010).

\bibitem{Riess:2020fzl} A.G.~Riess {\em et al.},  
Astrophys. J. Lett. \textbf{908}, L6 (2021).

\bibitem{Planck:2018vyg} N.~Aghanim \textit{et al.} [Planck],
Astron. Astrophys. \textbf{641}, A6 (2020).



\end{thebibliography}
\end{document}